\documentclass[epj]{svjour}

\usepackage{graphicx}
\usepackage{fancyhdr}
\usepackage{cite}

\def\mytitle{Gravitino Dark Matter and Light Element Abundances} 
\def\myauthors{Vassilis Spanos}  
\def\mytype{Parallel}
\def\mysession{Cosmology and Astrophysics}

\def\mytitle{Gravitino Dark Matter and Light Elements Adundance} 
\def\myauthors{Vassilis Spanos}    
\def\mytype{Contributed Talk}    
\def\mysession{Cosmology and Astrophysics}

\def\beq{\begin{equation}}
\def\eeq{\end{equation}}

\def\ga{\mathrel{\raise.3ex\hbox{$>$\kern-.75em\lower1ex\hbox{$\sim$}}}}
\def\la{\mathrel{\raise.3ex\hbox{$<$\kern-.75em\lower1ex\hbox{$\sim$}}}}
\def\gev{{\rm \, Ge\kern-0.125em V}}
\def\tev{{\rm \, Te\kern-0.125em V}}
\def\gyr{{\rm \, G\kern-0.125em yr}}

\def\gappeq{\mathrel{\rlap {\raise.5ex\hbox{$>$}}
{\lower.5ex\hbox{$\sim$}}}}
\def\lappeq{\mathrel{\rlap{\raise.5ex\hbox{$<$}}
{\lower.5ex\hbox{$\sim$}}}}
\def\Toprel#1\over#2{\mathrel{\mathop{#2}\limits^{#1}}}

\def\stau{\widetilde \tau}

\def\m12{m_{1\!/2}}

\newcommand\iso[2]{\mbox{${}^{#2}${\rm #1}}}
\def\he#1{\iso{He}{#1}}
\def\be#1{\iso{Be}{#1}}
\def\li#1{\iso{Li}{#1}}
\def\b1#1{\iso{B}{1#1}}

\def\stau{\tilde{\tau}}

\def\grav{\widetilde{G}}
\def\qbar{\overline{q}}

\def\bea{\begin{eqnarray}}
\def\eea{\end{eqnarray}}

\def\beqar{\begin{eqnarray}}
\def\eeqar{\end{eqnarray}}

\begin{document}
\title{Gravitino Dark Matter and Light Element Adundances}
\subtitle{}
\author{Vassilis Spanos 
\thanks{\emph{Email:} vspanos@physics.upatras.gr}%
}                     
%
%
\institute{ Department of Physics, University of Patras, GR-26500 Patras, Greece 
}
%
\date{}
\abstract{We discuss the scenario where the gravitino is the lightest supersymmetric particle and the
long-lived next-to-lightest sparticle (NSP) is the neutralino or the stau, the charged
partner of the tau lepton.
In this case staus form bound states with
several nuclei,  affecting  the cosmological abundances of \li6 and
\li7 by enhancing nuclear rates that would otherwise be strongly
suppressed. We consider the effects of these enhanced rates on the final
abundances produced in Big-Bang nucleosynthesis (BBN), including
injections of both electromagnetic and hadronic energy during and after
BBN. 
We show that if the stau lifetime is longer
than $10^3-10^4$~s, the abundances of \li6 and \li7 are far in excess of
those allowed by observations. For shorter lifetimes of order $1000$~s, 
it appears that 
stau bound state effect could reduce the \li7 abundance from standard BBN
values while at the same time enhancing the \li6 abundance, creating  a region 
  where both lithium abundances match their
plateau values.
} 
\maketitle

\section{Introduction}
\label{intro}

The primordial Big-Bang
nucleosynthesis (BBN)  predictions for the light elements abundance
provide some of the most stringent constraints on
the decays of unstable massive particles during the early
Universe~\cite{holtmann, kkm, kohri, cefo, kkm2, kmy, Jedamzik:2006xz,cefos}. 
This is because the
astrophysical determinations of the abundances of deuterium (D) and \he4  agree
well with those predicted by homogeneous BBN calculations, and also the
baryon-to-photon ratio $\eta \equiv n_b/n_\gamma \propto \Omega_b h^2$ 
needed for the success of these
calculations~\cite{cfo1,bbn2} agrees very well with that
inferred~\cite{cfo2} from observations of the power spectrum of
fluctuations in the cosmic microwave background (CMB). 
However, it is still difficult to reconcile the BBN predictions for the
lithium isotope abundances with observational indications on the
primordial abundances.  The discovery of the ``Spite''
plateau~\cite{spite}, which demonstrates a near-independence of the \li7
abundance from the metallicity in Population-II stars, suggests a
primordial abundance in the range ${\rm \li7/H} \sim (1 - 2) \times
10^{-10} $ ~\cite{rbofn}, whereas standard BBN with the CMB value of
$\eta$ would predict ${\rm \li7/H} \sim 4 \times 10^{-10}
$~\cite{cfo1,bbn2}. In the case of $\li6$, the data \cite{li6obs} lie a
factor $\sim 1000$ above the BBN predictions~\cite{bbnli6}. 

The effects of hadronic injections due to late decays of the NSP during
BBN have also been studied extensively~\cite{kohri,kkm2,kmy,karsten,jed,stef,kkm3,EOV,SFT}. 
It has recently been pointed out that, if it has electric charge, the NSP
forms bound states with several nuclei \cite{maxim}.
Due to the large NSP mass ($m_{\rm nsp} \gg m_{\rm nucleon}$),
the Bohr radii of these bound states 
$\sim \alpha^{-1} m_{\rm nucleon}^{-1} \sim 1 \ {\rm fm}$
are of order the nuclear size. Consequently, 
nuclear reactions with nuclei in bound states are catalyzed,
due to partial screening of the Coulomb barrier~\cite{kota,kapling},  and
due to the opening of virtual photon channels in radiative capture
reactions. 

Here, we present results
from a new analysis~\cite{cefos} that includes the nuclear reactions induced by
hadronic and electromagnetic showers generated by late gravitational decays of the NSP,
together with the familiar network of
nuclear reactions used to calculate the primordial abundances of the light
elements Deuterium (D), \he3, \he4 and \li7.
In addition, we include the effects of the bound states
when the decaying particle is charged. 
  We find that for lifetimes $\tau < 10^3 - 10^4$~s, the
enhanced rates of \li6 and \li7 production, exclude gravitino dark matter
(GDM) with a stau NSP.  At smaller lifetimes, we see that it is the \li7
destruction rates which are enhanced, facilitating a solution to the Li
problems.

\section{NSP Decays and resulting showers during BBN}
\label{sec:decays}
In order to estimate the lifetime of the NSP, as well as the various branching ratios
and the resulting EM and HD spectra, one must calculate the partial widths 
of the dominant relevant decay channels of the NSP.
The decay products that yield EM energy obviously include directly-produced photons,
and also indirectly-produced photons, charged leptons (electrons \- and \- 
muons) which are produced via the secondary decays of gauge 
and Higgs bosons, as well as neutral pions ($\pi^0$).
Hadrons (nucleons and  mesons such as the $K_L^0$, $K^\pm$ and $\pi^\pm $) 
are usually produced through the secondary decays of gauge and Higgs bosons,
as well (for the mesons) as via the decays of the heavy $\tau$ lepton.
It is important to note that mesons decay before interacting with the 
hadronic background~\cite{kohri,SFT}. Hence they are irrelevant 
to the BBN processes and to our analysis, except via their decays into photons
and charged leptons. Therefore, the HD injections on which we focus our attention are those that 
produce nucleons, namely the decays via gauge and Higgs bosons and quark-antiquark pairs.

For the neutralino NSP $\chi$, we include the two-body decay
channels $\chi \to \grav \, H_i$ and $\chi \to \grav \, V$, where $H_i=h,H,A$ and $V=\gamma,Z$. 
In addition, we include here the dominant 
three-body decays  $\chi \to \grav \, \gamma^*  \to \grav \, q  \qbar$,  
 $\chi \to \grav \, \gamma^* \to \grav \, W^+ W^-$, 
$\chi \to \grav \, W^+ W^-$ and the corresponding interference terms.
For the $\stau$ NSP case,  the
lighter stau is predominantly right-handed, its interactions with $W$
bosons are very weak (suppressed by powers of $m_\tau$) and can be
ignored. The decay rate for the dominant two-body decay channel, namely
$\stau \to \grav \, \tau$, has been given in~\cite{eoss5}. However, this
decay channel {\em does not yield any nucleons}.  Therefore, one must
calculate some three-body decays of the $\stau$ to obtain any protons or
neutrons. The most relevant channels are $\stau \to \grav \, \tau^* \to
\grav \, Z \, \tau $, $\stau \to Z \, \stau^* \to \grav \, Z \, \tau $,
$\stau \to \tau \chi^* \to \grav \, Z \, \tau $ and $\stau \to \grav \, Z
\, \tau $~\cite{SFT}.

Having calculated the partial decay widths and
\-  branching \- ratios, we employ
the {\small PYTHIA} event generator~\cite{pythia} to model both the EM and
the HD spectra  of the NSP decays.  
These spectra and the fraction of the energy of the decaying particle that is
injected as EM energy are then used to calculate the light-element
abundances, as it is described  in~\cite{cefos}.

\section{Bound-State Effects}
\label{sec:bs}

One the other hand,
it has recently been pointed out~\cite{maxim} that the presence of a charged
particle, such as the stau, during BBN can alter the light-element abundances in a significant way
due to the formation nega\- tively\- -charged staus of  bound states (BS)  with charged nuclei.
The binding energies of these states are
$\alpha^2 Z_i^2 m_i/2$  $\approx 30 Z_i^2 A_i \ {\rm keV}$,
and the Bohr radii $\sim (\alpha Z_i m_i)$  $\sim 1 \ Z_i^{-1} A_i^{-1} \ {\rm fm}$.
For species such as \he4, \li7 and \be7,
these energy and length scales are close to those of nuclear
interactions, and it thus turns out that bound state formation
results in catalysis of nuclear rates via two mechanisms.

One immediate
consequence of the bound states is a reduction of
the Coulomb barrier for nuclear reactions,
due to partial screening by the stau.
Since Coulomb repulsion dominates the charged-particle rates,
all such rates are enhanced.  
An additional effect enhances radiative capture channels $A_2(A_1,\gamma)X$ by
introducing photonless final states in which the stau
carries off the reaction energy transmitted via virtual photon processes.
In particular, the $\he4 (d,\gamma)\li6$ reaction, which is suppressed
in standard BBN, is enhanced by many orders of magnitude by the
presence of the bound states.
As described in~\cite{maxim}, the virtual photon channel has a cross
section which is enhanced over that of the usual radiative capture
cross section by
$\frac{\sigma_{\rm CBBN}}{\sigma_{\rm SBBN}} \sim (a \omega)^{-n}$,
where $a$ is the BS Bohr radius
and $\omega = \lambda^{-1}$ is the photon energy.
The index $n$ depends on the type of transition multipole:
for (E1,E2) transitions, $n = (3,5)$.
To account for bound state effects, an accurate calculation of their abundance
is necessary. To do this we solve numerically  the Boltzmann equations (13) and (14)
from~\cite{kota}, that control these abundances.

\section{Results and Discussion}
\begin{figure}
\includegraphics[width=0.43\textwidth,height=0.4\textwidth,angle=0]{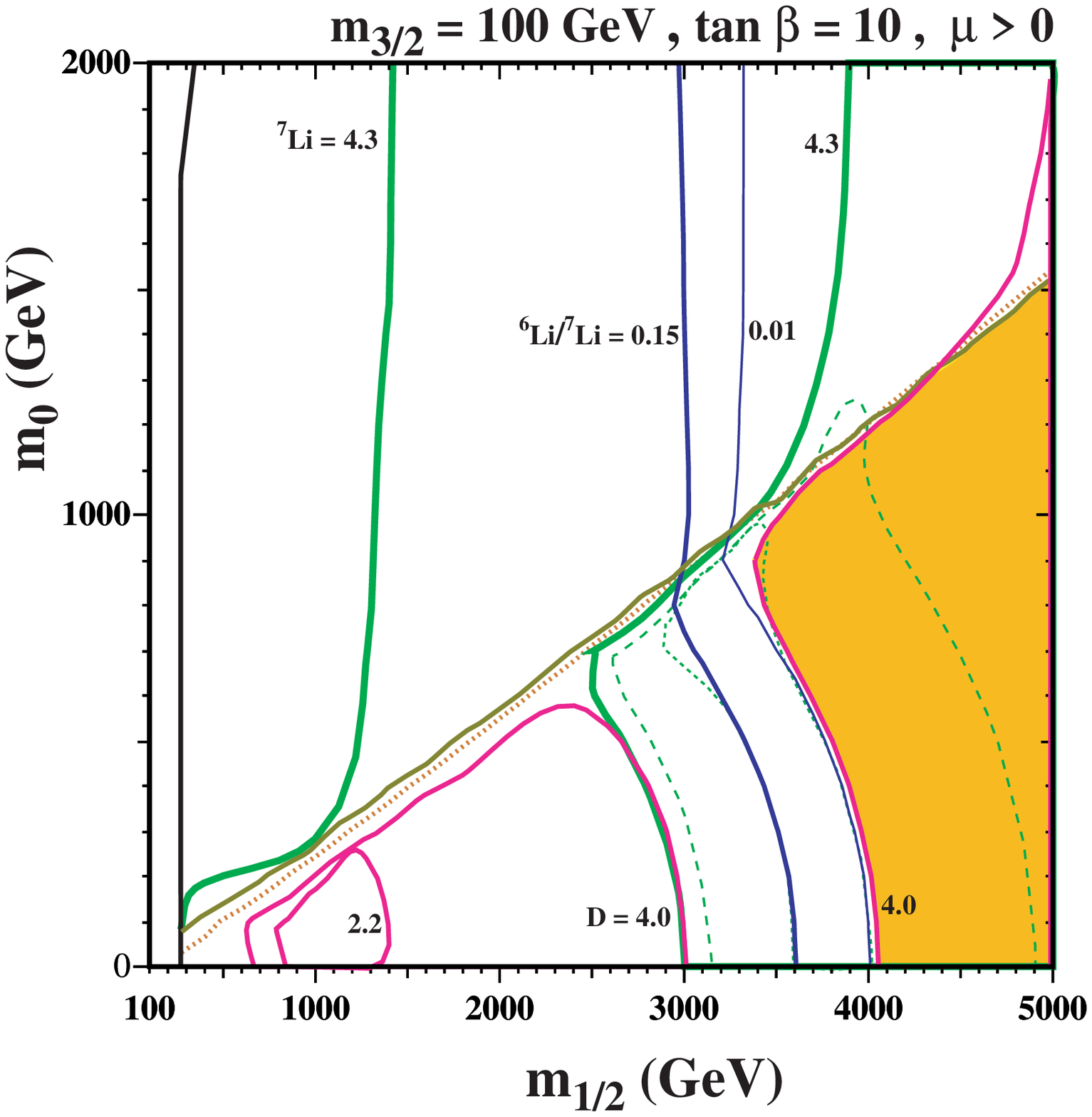} \\
\vspace*{0.5cm}
\includegraphics[width=0.43\textwidth,height=0.4\textwidth,angle=0]{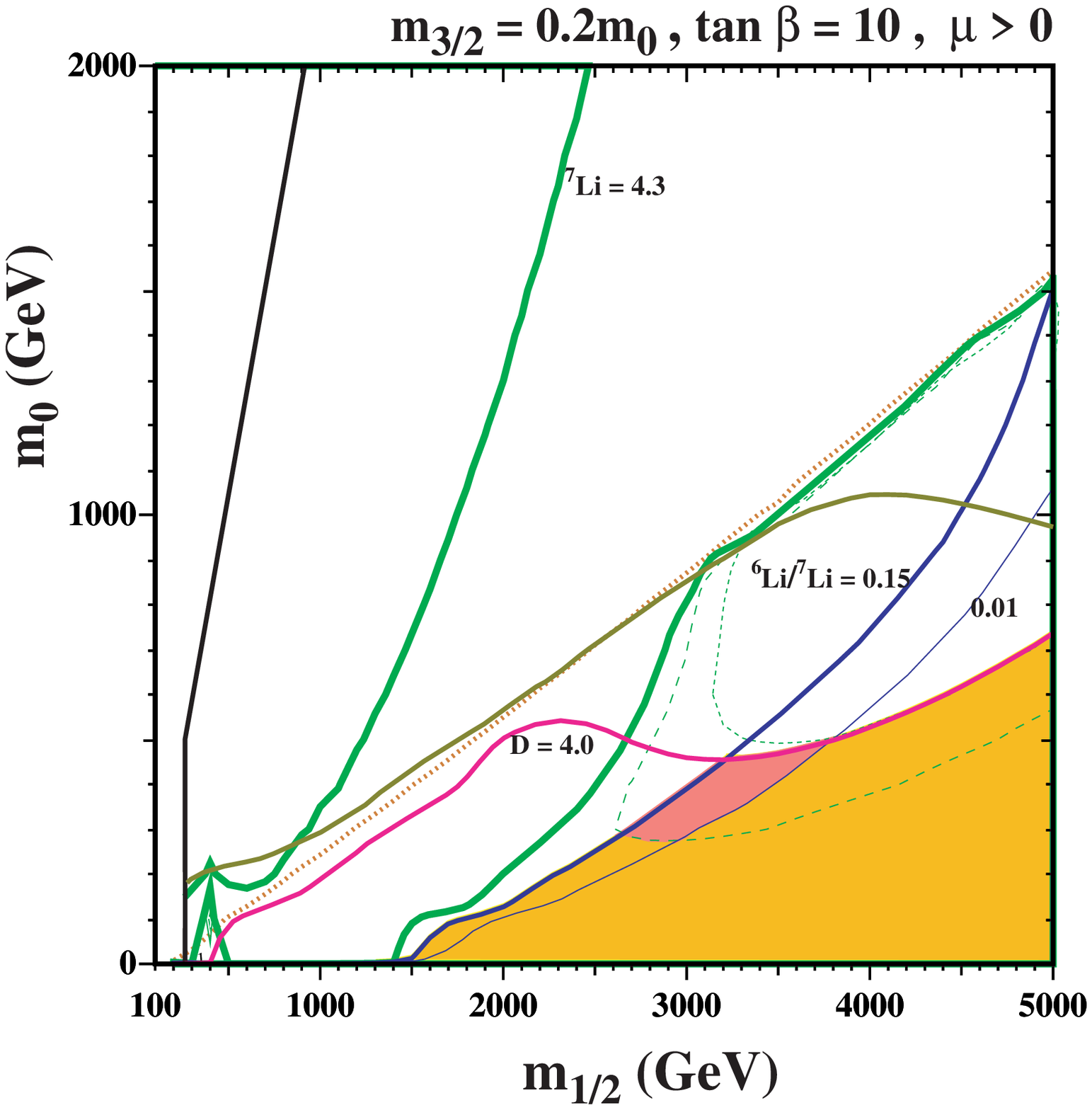}
\caption{
{\it 
The   $(m_{1/2}, m_0)$ planes for $A_0=0$, $\mu > 0$ and $\tan
\beta = 10$. In the upper (lower) panel we use $m_{3/2} = 100$~GeV
$(m_{3/2} = 0.2\, m_0)$.
The regions to the left of the solid black lines are not considered, 
since there the gravitino is not the LSP.
In the orange (light) shaded regions, the differences between the calculated 
and observed light-element abundances are no greater than in standard 
BBN without late particle decays. 
In the pink (dark) shaded region in the lower 
panel  the 
the lithium abundances match the observational plateau
values.
The significance of the 
other  contours  is explained in the text. In these figures we have  incorporated 
the bound-state effects. }}
\label{fig:fig1} 
\end{figure}

Our framework is  CMSSM and mSUGRA
models~\cite{vcmssm}, where the NSP could be either the lighter stau or
the lightest neutralino. 
We start presenting  results based on CMSSM models with $A_0 = 0$, $\mu > 0$ and
$\tan  \beta = 10$,
showing explicit element abundance contours.  
In Fig. \ref{fig:fig1}a, we show the element abundances that result
when the gravitino mass is held fixed at $m_{3/2} = 100$ GeV in the 
presence of stau bound state effects.
To the left of the near-vertical solid black line at $m_{1/2} \simeq 250$ GeV,
the gravitino is the not the LSP, and we do not consider this region here.
The diagonal red dotted line corresponds to the boundary between
a neutralino and stau NSP.  Above the line, the neutralino is the NSP,
and below it, the NSP is the stau.  Very close to this boundary,
there is a diagonal brown solid line.  Above this line, the relic
density of gravitinos from NSP decay is too high, i.e.,
$ (m_{3/2}/m_{NSP}) \Omega_{NSP} h^2 > 0.12$.
Thus we should restrict our attention to the area below this line.
Note that we display the extensions of contours which originate below the line into the overdense region, but we do not display 
contours that reside solely in the upper plane.

The very thick green line labelled \li7 = 4.3
corresponds to the contour where \li7/H = $4.3 \times 10^{-10}$, a value very
close to the standard BBN result for \li7/H. 
There are  additional (unlabeled) thin green contours showing \li7/H = 
$2 \times 10^{-10}$ (dashed). 
For this case with $m_{3/2} = 100$ GeV the \li6 abundance is never sufficiently
high to match the observed \li6 plateau for the same parameter values where \li7 is reduced.
The \li6/\li7 ratio is shown by the solid blue contour labeled \li6/\li7 = 0.15.
At large $m_{1/2}$, the contour for \li6/\li7 = 0.01 is shown by the thin blue line.
To the right of this contour, including the region where \li7 $\sim 2 \times 10^{-10}$,
the \li6 abundance is too small. 
Finally, we show the contours for D/H = 2.2 and 4.0 $\times 10^{-5}$
by the solid purple contours as labeled.  The D/H = 2.2  $\times 10^{-5}$
contour is a small loop within the \li6/\li7 loop.  Inside this loop
D/H is too small.  Between the two curves labeled 4.0, the D/H ratio is high, 
but not necessarily excessively so. 
As a better illustration of our results, we have shaded as
orange (lighter) the region  where  
the differences between the calculated and observed light- 
element abundances are no greater than in standard BBN 
without late particle decays, that is 
 $\he3/D < 1$,
$\li6/\li7 < 0.15$, $2.2\times 10^{-5} < D/\textrm{H} < 4.0 \times10^{-5}$ and $\li7/\textrm{H}
< 4.3 \times10^{-10} $. 
Only when $m_{1/2} \ga 3500 - 4000$ GeV does the D/H abundance
drop back to acceptable levels with good abundances for \li7, but \li6 is
now too small to account for the plateau.  Thus, for a constant value of
$m_{3/2} = 100$ GeV, the bound-state effects force one to extremely large
values of $m_{1/2}$ primarily due to the enhanced production of \li6, as
shown by the orange shaded region. {\it For this value of the gravitino
mass, there are no regions where both lithium abundances match their
plateau values.}

 In Fig.~\ref{fig:fig1}b, we fix $m_{3/2} = 0.2\, m_0$ and neglect 
 the bound-state effects. The choices of contours are
similar to the upper panel.  The gravitino relic density
constraint now cuts out some of the stau NSP region at large $m_{1/2}$ and
large $m_0$, but allows a small neutralino NSP region at low $m_{1/2}$. 
In this case the constraint from \he3/D is not very strong in the stau NSP
region and the contour is not shown.  
The region where the
\li6/\li7 ratio lies between 0.01 and 0.15 now forms a band which moves
from lower left to upper right.  Thus, as one can see in the orange
shading, there is a large region where the lithium isotopic ratio can be
made acceptable. However, if we restrict to D/H $< 4.0 \times 10^{-5}$, we
see that this ratio is interesting only when \li7 is at or slightly below
the standard BBN result. 
Once again, we see that the increased production
of both \li6 and \li7 excludes a portion of the stau NSP region where
$m_{1/2} \la 1500$ GeV for small $m_0$.  The lower bound on $m_{1/2}$
increases with $m_0$.  In this case, not only do the bound-state effects
increase the \li7 abundance when $m_{1/2}$ is small (i.e., at relatively
long stau lifetimes), but they also decrease the \li7 abundance when the
lifetime of the stau is about 1500~s. Thus, at $(m_{1/2}, m_0) \simeq
(3200,400)$, we find that \li6/\li7 $\simeq 0.04$, \li7/H $\simeq 1.2
\times 10^{-10}$, and D/H $\simeq 3.8 \times 10^{5}$.  Indeed, when
$m_{1/2}$ is between 3000-4000 GeV, the bound state effects cut the \li7
abundance roughly in half. {\it In the darker (pink) region 
the lithium abundances match the observational plateau
values}, with the properties $\li6/\li7 > 0.01$ and $0.9 \times10^{-10} <
\li7/\textrm{H} < 2.0 \times10^{-10}$.

\begin{figure}
\includegraphics[width=0.43\textwidth,height=0.4\textwidth,angle=0]{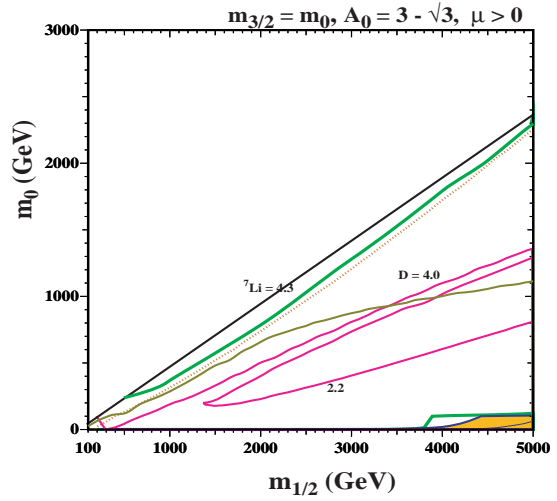} \\
\caption{
{\it 
The  $(m_{1/2}, m_0)$ plane for mSUGRA with $m_{3/2} = m_0$ and $A_0/m_0=3-\sqrt{3}$ as
in the simplest Polonyi superpotential.
The various contours and regions are as in Fig.~\ref{fig:fig1}. }}
\label{fig:fig2} 
\end{figure}

Finally, we come to an example of a mSUGRA model in Fig.~\ref{fig:fig2}. 
 Here, because of a
relation between the bilinear and trilinear supersymmetry breaking terms:
$B_0 = A_0 - m_0$, $\tan \beta$ is no longer a free parameter of the
theory, but instead must be calculated at each point of the parameter
space.  Here, we choose an example based on the Polonyi model for which
$A_0/m_0 = 3- \sqrt{3}$.  In addition, we have  $m_{3/2}
= m_0$.  
The upper part of the plane, we do not have GDM. 
We note that \li6 is interestingly high,
between 0.01 and 0.15 in much of this region.
Due to the bound-state effects  both lithium isotope abundances are too large except
in the extreme lower right corner, where there is a small region shaded 
orange. Henceforth,  the BBN constraints and the bound state effects practically exclude 
the bulk of the parameter space of 
this  simple  Polonyi SUGRA model.

\section{Conclusions}

We discussed  the cosmological light-element \- abundances
in the presence of the electromagnetic and \- hadronic showers due to late
decays of the NSP in the context of the CMSSM and mSUGRA models,
incorporating the effects of the bound states that would form between a
metastable stau NSP and the light nuclei.  Late decays of the neutralino
NSP constrain significantly the neutralino region, since in general they
yield large light-element abundances.  The bound-state effects are
significant in the stau NSP region, where excessive \li6 and \li7
abundances exclude regions where the stau lifetime is longer than
$10^3-10^4$~s. For lifetimes shorter than $1000$~s, there is a possibility
that the stau decays can reduce the \li7 abundance from the standard BBN
value, while at the same time enhancing the \li6 abundance,
defining   a region 
  where both lithium abundances match their
plateau values. 
 
\section*{Acknowledgments}
\noindent 
This work  was supported by Marie Curie International Reintegration grant
 ``SUSYDM-PHEN",  MIRG-CT-2007-203189 and the 
 Marie Curie Excellence grant  \-  MEXT-CT-2004-014297. We also 
acknowledge support from the Research Training Network 
``HEPTOOLS",  MRTN-CT-2006-035505.


\end{document}